\documentclass[twocolumn,showpacs,preprintnumbers,amsmath,amssymb]{revtex4}

\usepackage{graphicx}
\usepackage{dcolumn}
\usepackage{bm}

\begin{document}
\title{Electron dephasing near zero temperature: an experimental review}
\author{J. J. Lin}
\address{Institute of Physics, National Chiao Tung University, Hsinchu 300, Taiwan}

\author{T. J. Li}
\address{Department of Electrophysics, National Chiao Tung University, Hsinchu 300, Taiwan}

\author{Y. L. Zhong}
\address{Department of Physics, National Tsing Hua University, Hsinchu 300, Taiwan}

\begin{abstract}

The behavior of the electron dephasing time near zero temperature, $\tau_\phi^0$, has recently
attracted vigorous attention. This renewed interest is primarily concerned with whether
$\tau_\phi^0$ should reach a finite or an infinite value as $T \rightarrow$ 0. While it is
accepted that $\tau_\phi^0$ should diverge if there exists only electron-electron
(electron-phonon) scattering, several recent measurements have found that $\tau_\phi^0$ depends
only very weakly on temperature, if at all, when $T$ is sufficiently low. This article discusses
the current experimental status of ``the saturation problem", and concludes that the origin(s) for
this widely observed saturation are still unresolved.

\end{abstract}

\maketitle

\section{Introduction}

The electron dephasing time, $\tau_\phi$, is one of the most important quantities governing
quantum-interference phenomena in mesoscopic structures. Recently, the behavior of the dephasing
time near zero temperature, $\tau_\phi^0 = \tau_\phi (T \rightarrow 0)$, has attracted vigorous
experimental \cite{Webb97a,Birge00,Lin01a,Natelson01,Ovadyahu01,Bird99c,Marcus99} and theoretical
\cite{Alt98,Zaikin98,Zawa99,Imry99,Kaminski01,Imry02} attention. One of the central themes of this
renewed interest is concerned with whether $\tau_\phi^0$ should reach a finite or an infinite
value as $T \rightarrow$ 0. The connection of the $\tau_\phi^0$ behavior with fundamental
condensed-matter physics problems, such as the validity of the Fermi-liquid picture
\cite{Webb98b}, the possibility of the occurrence of a quantum phase transition, the persistent
current problem in metals \cite{Mohanty99,persist}, and also the feasibility of quantum computing,
has been intensively addressed \cite{Mohanty00}. Conventionally, it is accepted that $\tau_\phi^0$
should reach an infinite value in the presence of only Nyquist electron-electron (e-e) and
electron-phonon (e-ph) scattering. However, several recent careful measurements, performed on
metal and semiconductor mesoscopic structures, have revealed that $\tau_\phi^0$ depends only very
weakly on temperature, if at all, when $T$ is sufficiently low. Until now, there is {\em no
generally} accepted process of electron--low-energy-excitation interaction that can satisfactorily
explain the widely observed saturation of $\tau_\phi^0$.

This article discusses existing proposals for the observed saturation of $\tau_\phi^0$ and surveys
recent systematic efforts aimed at testing these proposals. We argue that recent measurements have
extensively demonstrated electron heating, external microwave noise, and very dilute magnetic
impurities {\em cannot} be the dominant source for the finite value of $\tau_\phi^0$ found in the
experiments. We suggest that ``the saturation problem" can be most unambiguously addressed using
tailor-made samples covering a wide range of material properties. We also propose that
three-dimensional (3D) mesoscopic structures can shed light on this issue.

\section{Extracting $\tau_\phi$ from magneto-transport measurements}

In quantum-interference studies, the electron dephasing time $\tau_\phi$ is given by \cite{Alt85}
\begin{equation}
\frac{1}{\tau_\phi (T,\ell)} = \frac{1}{\tau_\phi^0 (\ell)} + \frac{1}{\tau_{\rm i}(T,\ell)} \,,
\label{eq1}
\end{equation}
where $\ell$ is the electron elastic mean free path, $\tau_\phi^0$ is presumed to be independent
of temperature, and $\tau_{\rm i}$ is the relevant inelastic electron scattering time(s) in
question. The temperature dependence of $\tau_\phi$ is controlled entirely by the $T$ dependence
of $\tau_{\rm i}$, while $\tau_\phi^0$ determined the value of the dephasing time in the limit of
very low $T$. In three dimensions (3D), the e-ph scattering is the {\em sole}, dominating
inelastic scattering \cite{Lin02r,Gershen99} and $\tau_{\rm i}^{-1} \approx \tau_{ep}^{-1} \propto
T^p$, with $2 \lesssim p \lesssim 4$. In lower dimensions, the Nyquist e-e scattering dominates
\cite{Alt85} and, generally, $\tau_{\rm i}^{-1} \approx \tau_{ee}^{-1} \propto T^p$ at a few
kelvins and lower, with $p$ = 2/3 in one dimension (1D) and $p$ = 1 in two dimensions (2D).

\begin{figure}
\includegraphics[scale=0.28,angle=270]{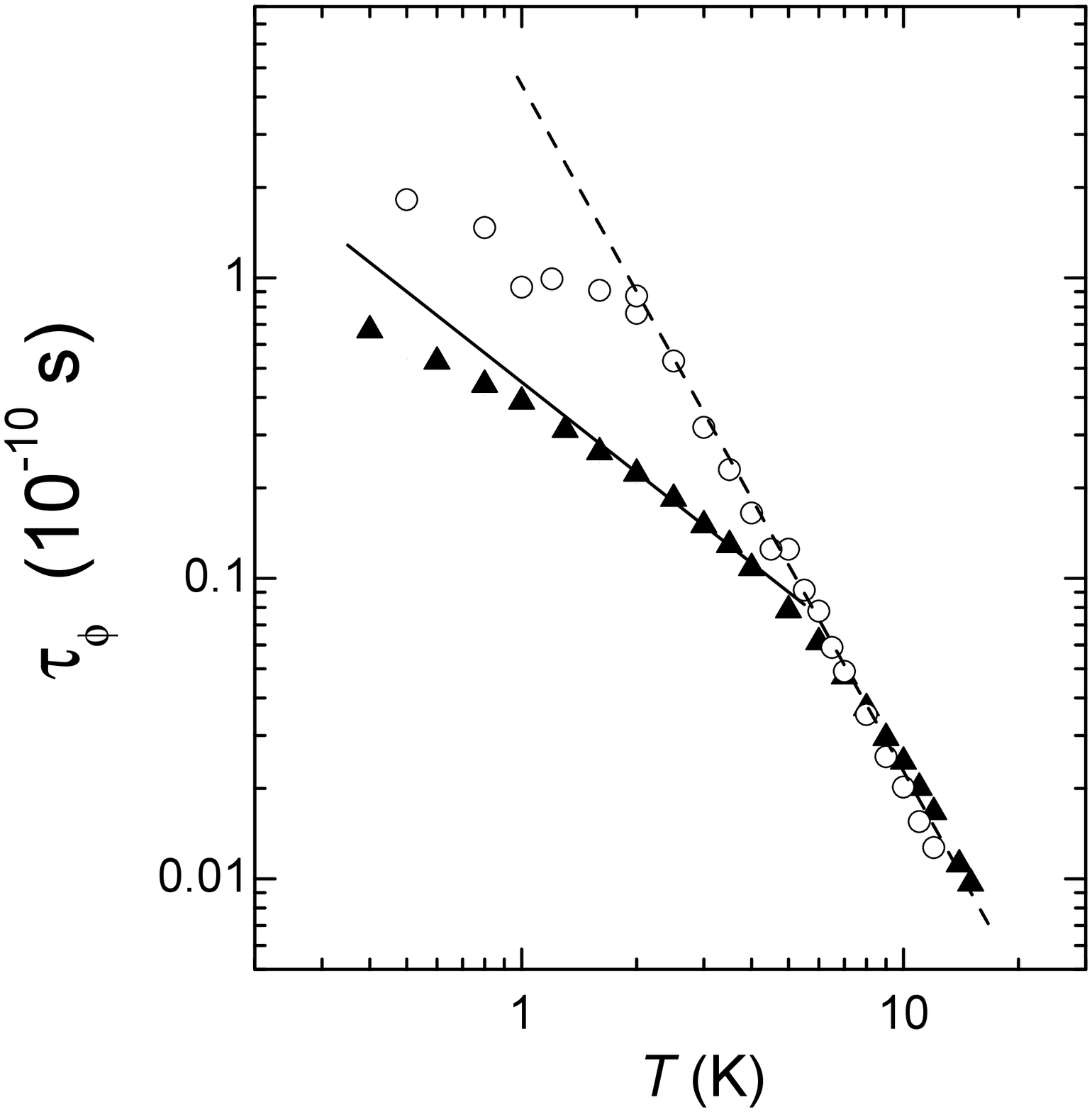}
\caption{Variation of $\tau_\phi$ with temperature for a 3000-$\rm\AA$ (circles) and a
175-$\rm\AA$ (triangles) thick Sb films. The thick and thin films have resistivities $\rho$(10\,K)
= 1320 and 1920 $\mu \Omega$ cm, respectively. The dashed line is a guide to the eye and the solid
line is drawn proportional to $T^{-1}$. \label{f1} }
\end{figure}

It is well established that {\em very reliable} values of $\tau_\phi$ in mesoscopic structures can
now be extracted from magneto-transport measurements \cite{Lin02r}. For example, Fig. \ref{f1}
shows the variation of $\tau_\phi$ with $T$ for a thick and a thin Sb films obtained from
weak-localization (WL) studies. One sees that, at the highest measurement temperatures, e-ph
scattering dominates and $\tau_\phi$ reveals a strong $T$ dependence. In the thick film,
$\tau_\phi^{-1} \approx \tau_{ep}^{-1}$ all the way down to about 2 K; while in the thin film, e-e
scattering gradually becomes more important and, thus, $\tau_\phi^{-1} \approx \tau_{ee}^{-1}$
below about 5$-$6 K. {\em In both cases, a progressively weakened $T$ dependence is observed at
the lowest measurement temperatures.} We notice that (i) the overlap of the values of $\tau_{ep}$
for the two films at high measurement temperatures, and (ii) the crossover of $\tau_\phi$ to e-e
scattering in the thin film (but not in the thick film!) at lower temperatures, provide a
convincing consistency check for the experimental $\tau_\phi$ inferred from WL studies.

\begin{figure}
\begin{center}
\includegraphics[scale=0.28,angle=270]{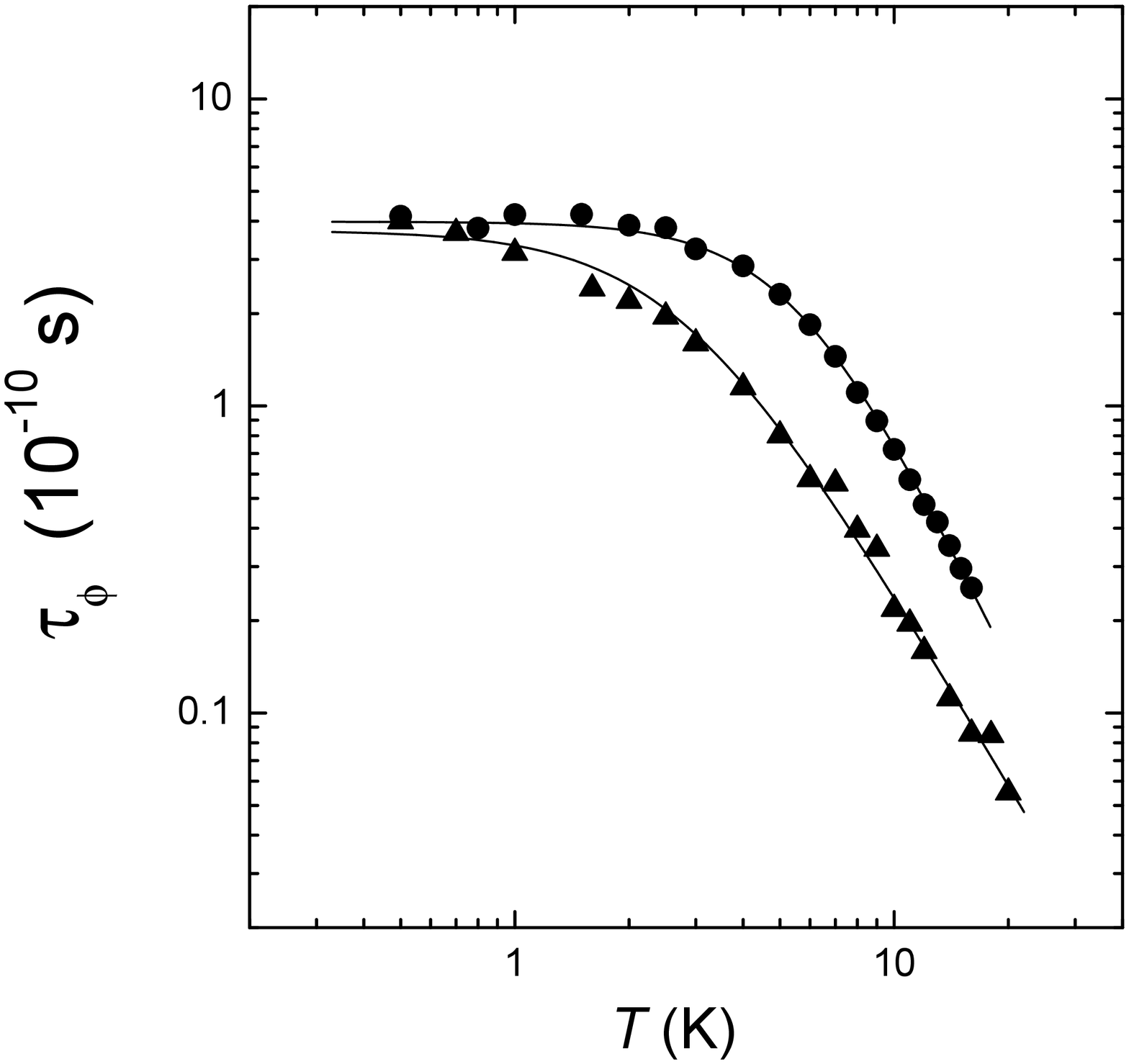}
\caption{Variation of $\tau_\phi$ with temperature for a DC- (triangles) and a RF-sputtered
(circles) Pd$_{60}$Ag$_{40}$ thick films. The two films have $\rho$(10\,K) = 180 $\mu \Omega$ cm.
The solid curves are guides to the eye. \label{f2} }
\end{center}
\end{figure}

Figure \ref{f2} shows the variation of $\tau_\phi$ with temperature for a DC- and a RF-sputtered
Pd$_{60}$Ag$_{40}$ thick films obtained from WL studies. These two thick films are 3D and have a
similar level of disorder. Notice that they assumes a similar value of $\tau_\phi^0$ as $T
\rightarrow 0$, independent of the different fabrication methods. This observation provides an
alternative convincing consistency check for the experimental $\tau_\phi$ extracted. (See further
discussion below.)

\section{Electron heating and related effects}

Of the prominent interest on this subject is ``the saturation problem", i.e., a finite
experimental value of $\tau_\phi^0$ extracted as $T \rightarrow 0$. In particular, in the WL
studies, the saturation of $\tau_\phi^0$ is inferred from the low-field magnetoresistance, which
does not increase as fast as expected with decreasing $T$ \cite{Natelson01,Lin02r}. For instance,
Natelson {\it et al.} \cite{Natelson01} found, in wide AuPd wires, that the magnetoresistance
changed by less than 40\% even when the temperature was significantly decreased from 4.2 down to
0.08 K. On the other hand, it should be noted that the temperature insensitive magnetoresistance
at low $T$ is found in a $T$ regime where the electron gas is {\em in thermal equilibrium} with
the lattice, i.e. the weak dependence is {\em not} caused by electron heating (due to either the
measurement current or external noise). This assertion is clearly confirmed by the observation of
a continuous resistance rise with decreasing temperature down to the lowest temperatures in each
measurement \cite{Natelson01,Webb97a,Lin01a,Ikoma96}. (The resistance rise can generally be
ascribed to e-e interaction effects in metal samples \cite{Alt85,Fuku85}.) This assertion is also
confirmed in other quantum-interference studies. For example, Bird {\it et al.} \cite{Bird90} have
extracted estimates for the dephasing length, $L_\phi = \sqrt{D \tau_\phi}$, where $D$ is the
diffusion constant, in quasi-ballistic GaAs/AlGaAs quantum wires from the amplitude of the
universal conductance fluctuations. They found that $L_\phi$ remained independent of $T$ below 1
K, even though the amplitude of the fluctuations themselves increased by a factor of 4 over the
same range \cite{they}.

{\it Microwave-noise dephasing.} Altshuler {\it et al.} \cite{Alt81b} have considered the electron
dephasing by non-equilibrium high-frequency electromagnetic noise. They have argued that the
microwave noise can already be large enough to cause dephasing, while still too small to cause
significant Joule heating of the conduction electrons \cite{Alt98,Gershen99}. Careful experimental
measurements have recently been designed to test these predictions. These experiments
\cite{Marcus99,Webb98a,Burke02} explicitly demonstrated that direct dephasing due to radiation
could {\em not} be the cause of the widely observed saturation. More precisely, Webb {\it et al.}
\cite{Webb98a} and Huibers {\it et al.} \cite{Marcus99} found that there was heating by the
high-frequency noise, before it affected dephasing, i.e. electron heating {\em preceded} dephasing
by high-frequency noise. Burke {\it et al.} \cite{Burke02} have very recently investigated the
effect of externally applied {\em broadband} Nyquist noise on the intrinsic dephasing rate of
electrons in 2D GaAs/AlGaAs heterojunctions at low temperatures. They also found {\em no major}
change in the measured $\tau_\phi$ even when their sample was subject to {\em large-amplitude},
externally-applied, voltage fluctuations. (At the same time, heating was unimportant in their
measurements.) These measurements, therefore, strongly suggests that the effect of microwave noise
on electron dephasing and heating requires further theoretical clarification.

\section{Magnetic impurities: spin-spin scattering}

Over the years, the saturation behavior of $\tau_\phi^0$ has often been ascribed to a finite
spin-spin scattering rate, due to the presence of a tiny amount of magnetic impurities in the
sample. Such a finite scattering rate will eventually dominate over the relevant inelastic
scattering in the limit of sufficiently low $T$, Eq. (\ref{eq1}). This idea of
magnetic-scattering-induced dephasing immediately became widely accepted since the discovery of WL
effects two decades ago \cite{Hikami80}. In addition to many early studies that often attributed
the observed finite value of $\tau_\phi^0$ to spin-spin scattering, there are some recent studies
that also argue in favor of the role of magnetic impurities. Especially, the Saclay-MSU group
\cite{Birge00} has measured both the energy exchange rate between quasiparticles and the dephasing
time of quasiparticles in several Cu, Ag, and Au narrow wires. They found in one Ag wire and one
Au wire that $\tau_\phi$ varies as $T^{-2/3}$ down to 40 mK. (The $T^{-2/3}$ variation is expected
from 1D Nyquist e-e scattering \cite{Alt85,Fuku85}.) Comparing these two complementary
measurements, they concluded that a saturation of $\tau_\phi$ occurs only in wires that contain a
small amount of magnetic impurities. In those wires where they found no anomalous energy exchange,
they also found {\em no} sign of saturation in $\tau_\phi^0$. The Saclay-MSU group experimental
results have triggered several theoretical studies \cite{Kaminski01,Goppert01,Kroha02} of the
inference of one-channel and two-channel Kondo effects on the energy-relaxation and dephasing
rates. We notice that the metal wires studied by the Saclay-MSU group are relatively ``clean",
namely, their wires have a diffusion constant $D \approx$ 100$-$200 cm$^2$/s.

{\it Proposal for non-magnetic origin.} In sharp contrast to the conclusion reached by the
Saclay-MSU group discussed just above, Mohanty {\it et al.} \cite{Webb97a} have tested and argued
for a {\em non-magnetic} origin for the finite value of $\tau_\phi^0$. Mohanty {\it et al.} first
studied a series of very pure Au wires, finding that there was always a saturation of
$\tau_\phi^0$. From these measurements, they realized that both the value of $\tau_\phi^0$ and the
onset temperature of saturation could be tuned by adjusting the sample parameters such as the wire
length, resistance, and diffusion constant. To explore this idea, Webb {\it et al.}
\cite{Webb98a,Webb98c} reported further measurements on several carefully fabricated Au wires and
films, whose onset temperature of saturation was indeed pushed down to unattainable temperatures
($\ll$ 40 mK). Webb {\it et al.} also noticed that a finite value of $\tau_\phi^0$ is also often
observed in semiconductor mesoscopic structures. Since such structures are thought to contain only
the smallest concentration of magnetic impurities, they concluded that the widely observed
saturation must be universal, and {\em cannot} be simply due to magnetic scattering. (They
suggested that the saturation of $\tau_\phi^0$ is intrinsic and is a signature of the breakdown of
the independent single-electron picture in mesoscopic systems as $T \rightarrow 0$.)

The contradicting conclusion of the Saclay-MSU group and Webb {\it et al.} illustrates well the
subtlety and complexity of ``the saturation problem". First, it is {\em not} a trivial
experimental task to unambiguously determine the influence of magnetic scattering on
$\tau_\phi^0$, because the level of magnetic contamination is probably so low that it cannot be
readily detected with state-of-the-art material-analysis techniques. Secondly, since there are no
known physical properties that are more sensitive to spin-flip scattering than the dephasing
process, the problem of whether there is a tiny amount of magnetic contamination in the sample,
thus, cannot be readily verified with other complimentary measurements. Moreover, the situation
becomes even more serious when {\em lower-dimensional} systems are considered. In the case of
low-dimensional structures, surface effects due to interfaces, substrates, and paramagnetic
oxidation \cite{Haesendonck88} are likely to be non-negligible. Then, it is not straightforward to
ascribe the observed saturation behavior of $\tau_\phi^0$ to either intrinsic material properties
or surface effects.

\section{Systematic measurements and the importance of three-dimensional systems}

To resolve the underlying physics of $\tau_\phi^0$, the usual experimental approach of measuring
the inelastic electron processes via temperature-dependent magnetoresistance studies is not very
useful. In the case of inelastic scattering, the microscopic physics of the relevant
electron--low-energy-excitation interactions is extracted through the measured variation of the
scattering time with $T$. However, in the case of $\tau_\phi^0$, there is only a very weak, or no,
$T$ dependence involved. It is then desirable to seek variations of $\tau_\phi^0$ with the
material characteristics of the samples, such as the amount of disorder \cite{Lin01a}, the sample
geometry \cite{Natelson01}, the effect of annealing \cite{Lin02,Lin87b}, and the effect of the
microscopic quality of disorder \cite{Ovadyahu01,Lin02r}. Systematic information about the
influence of sample properties on $\tau_\phi^0$ should shed light on the origins of the zero-$T$
dephasing mechanism.

As discussed above, an explanation for the saturation behavior of $\tau_\phi^0$ based on magnetic
scattering cannot be easily discerned experimentally. This experimental difficulty results in
several groups insisting on the presence of magnetic impurities in the sample as the origin of
saturation. In our opinion, this problem may be resolved by studying a series of samples covering
a {\em sufficiently wide} range of sample properties. For instance, Lin and Giordano \cite{Lin87b}
have performed systematic measurements of $\tau_\phi^0$ on a number of as-sputtered and annealed
AuPd thin films with varying sheet resistance. They found a $\tau_\phi^0$ increasing with
decreasing sheet resistance, which led them to conclude that magnetic scattering could {\em not}
be the mechanism responsible. They suggested that the strength of the impurity scattering which is
responsible for $\tau_\phi^0$ could be very sensitive to the {\em metallurgical properties} of the
films, which are in turn a function of both thickness and annealing, etc. Since two-level systems
(TLS) are closely associated with the presence of dynamical defects in the microstructures in the
sample, their observation of a sensitive, metallurgical-property, influence on $\tau_\phi^0$ has
recently inspired several theoretical studies of the interaction between conduction electrons and
TLS \cite{Zawa99,Imry99}.

\subsection{Three-dimensional polycrystalline metals}

\begin{figure}
\begin{center}
\includegraphics[scale=0.3,angle=270]{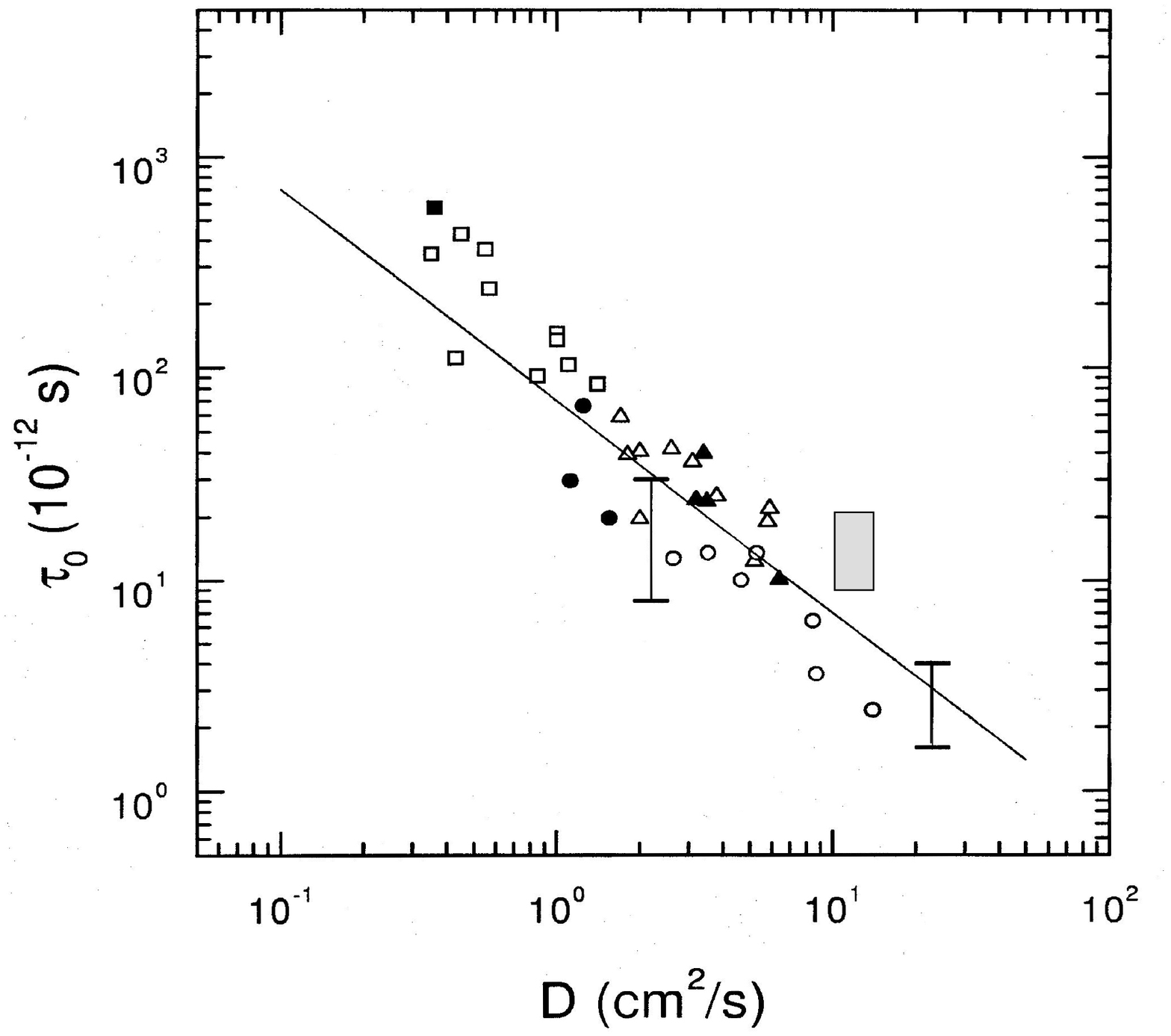}
\caption{Variation of $\tau_\phi^0$ with diffusion constant for 3D polycrystalline metals: dc
sputtered Au$_{50}$Pd$_{50}$ (circles), dc/RF sputtered Pd$_{60}$Ag$_{40}$ (squares), dc sputtered
Sb (triangles), thermal-flash evaporated Au$_x$Al (solid triangles), thermal-flash evaporated
Sc$_{85}$Ag$_{15}$ (solid squares), and arc-melted V$_{100-x}$Al$_x$ (solid circles). The two
vertical bars represent the $\tau_\phi^0$ measured in AuPd thin films in Ref. \cite{Lin87b}. The
shaded area represents the $\tau_\phi^0$ measured in AuPd wires and films in Ref.
\cite{Natelson01}. (Notice that these AuPd films and wires have short electron mean free paths and
are 3D with regard to the Boltzmann transport.) The solid line is drawn proportional to $D^{-1}$
and is a guide to the eye. \label{sf7} }
\end{center}
\end{figure}

Lin and Kao \cite{Lin01a} have recently studied the electron dephasing times $\tau_\phi$ in
numerous 3D {\em polycrystalline} disordered metals. Their samples were made of various materials,
using various fabrication techniques (see the caption to Fig. \ref{sf7}). Since one of the major
issues in this direction of research is to study whether there might exist a universal saturation
behavior of $\tau_\phi^0$, the use of many kinds of samples with distinct characteristics is
highly desirable. Any behavior of $\tau_\phi^0$ common to all these materials, if found, should
bear important information on the nature of the zero-$T$ dephasing. Regardless of the very
different preparation and measurement conditions, the authors found in numerous metals that there
is a saturation of $\tau_\phi$ at sufficiently low $T$. Most surprisingly, they found that their
experimental $\tau_\phi^0$ varied with the diffusion constant with a simple power law as
\begin{equation}
\tau_\phi^0 \propto D^{- \alpha} \,, \,\,\,\,\,\, \alpha \gtrsim 1 \label{4ss}
\end{equation}
where $\alpha$ is {\em close to} or {\em slightly larger} than 1.

Figure~\ref{sf7} shows the variation of $\tau_\phi^0$ with $D$ measured by Lin and Kao. This
figure indicates that the values of $\tau_\phi^0$ for all samples fall essentially on a universal
dependence. Particularly, it reveals that, regardless of the distinct material characteristics
(e.g., electronic structure) of the various samples, all that matters in determining the value of
$\tau_\phi^0$ is $D$. (Figure~\ref{f2} is a straight manifestation of this observation.) This
observation of $\tau_\phi^0 \propto D^{- \alpha}$, with $\alpha \gtrsim 1$, is totally unexpected.
This result implies that the functional form of $\tau_\phi^0$ on disorder may be universal for a
given dimensionality and a given kind of sample structure, while it may {\em not} be universal
over different dimensionalities and different sample (e.g., polycrystalline, amorphous or
well-textured semiconductor) structures. On the contrary, it is often conjectured that
$\tau_\phi^0$ should increase with {\em reducing} disorder, {\em at least} in lower dimensions
\cite{Webb97a,Ikoma96}. Until now, it is {\em not} known exactly how differently $\tau_\phi^0$
should behave in different dimensionalities and in different sample structures. This observation
may also suggest that the saturation behavior of $\tau_\phi^0$ can be very different between
``clean" and ``dirty" metals. For comparison, the diffusion constant considered in Fig. \ref{sf7}
is typically 1 to 2 orders of magnitude smaller than that in the metal wires studied by the
Saclay-MSU group \cite{Birge00} and Mohanty {\it et al}. \cite{Webb97a,Webb98a}. On the other
hand, the diffusion constant in the AuPd wires and films studied by Natelson {\it et al.}
\cite{Natelson01} is similar to that considered in Fig. \ref{sf7}. Consequently, Natelson {\it et
al.} obtained corresponding values of $\tau_\phi$ very close to that shown in Fig. \ref{sf7}.

The result of Fig. \ref{sf7} argues against the role of magnetic scattering as the dominant
dephasing process in 3D polycrystalline metals as $T \rightarrow 0$. This is asserted since the
numerous samples considered in Fig. \ref{sf7} were made from very different high-purity sources,
using very different fabrication techniques. It is hard to conceive that spin-flip scattering due
to ``unintentional" magnetic contamination could have caused the ``systematic" variation given by
Eq. (\ref{4ss}). If magnetic scattering were responsible for the measured $\tau_\phi^0$ in Fig.
\ref{sf7}, then the unintentional magnetic impurity concentration, $n_m$, must vary {\em randomly}
from sample to sample, and hence one should expect a {\em random} $\tau_\phi^0$ ($\propto
n_m^{-1}$), independent of disorder. Besides, any spin-spin scattering that might result from
surface effects (substrates, paramagnetic surface oxidation, etc.) should be largely minimized in
these 3D samples. Therefore, the result of Fig. \ref{sf7} {\em cannot} be simply explained in
terms of magnetic scattering.

The observation of Fig. \ref{sf7} is still not understood. Nevertheless, this result unambiguously
indicates that the saturation of $\tau_\phi^0$ in this case is certainly {\em not} due to
microwave noises, because microwave-noise dephasing should result in a $\tau_\phi^0 \propto
D^{-1/3}$ dependence in 3D \cite{Alt81b}.

\subsection{Effect of annealing: 3D polycrystalline metals}

\begin{figure}
\begin{center}
\includegraphics[scale=0.28,angle=270]{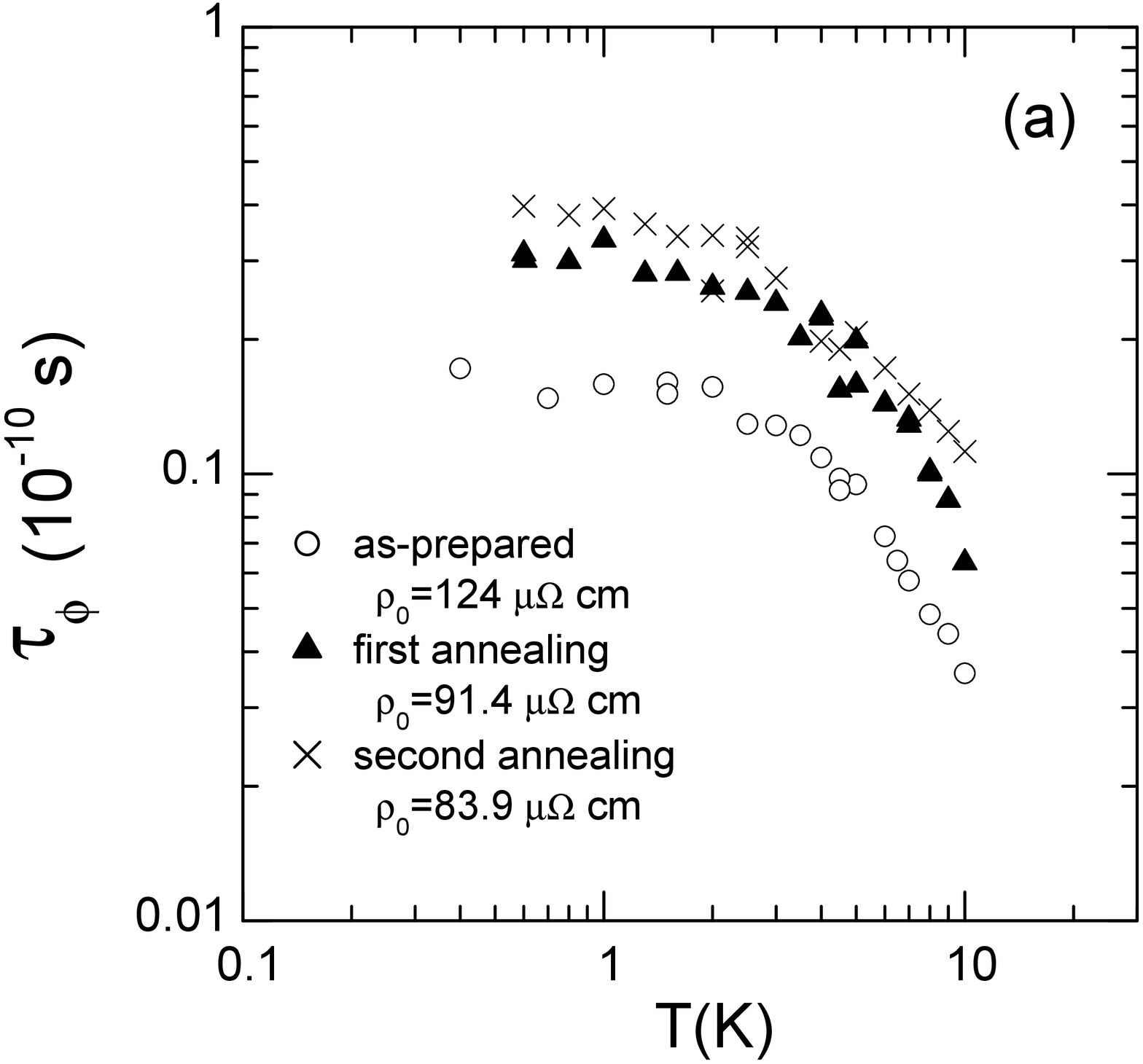}
\includegraphics[scale=0.28,angle=270]{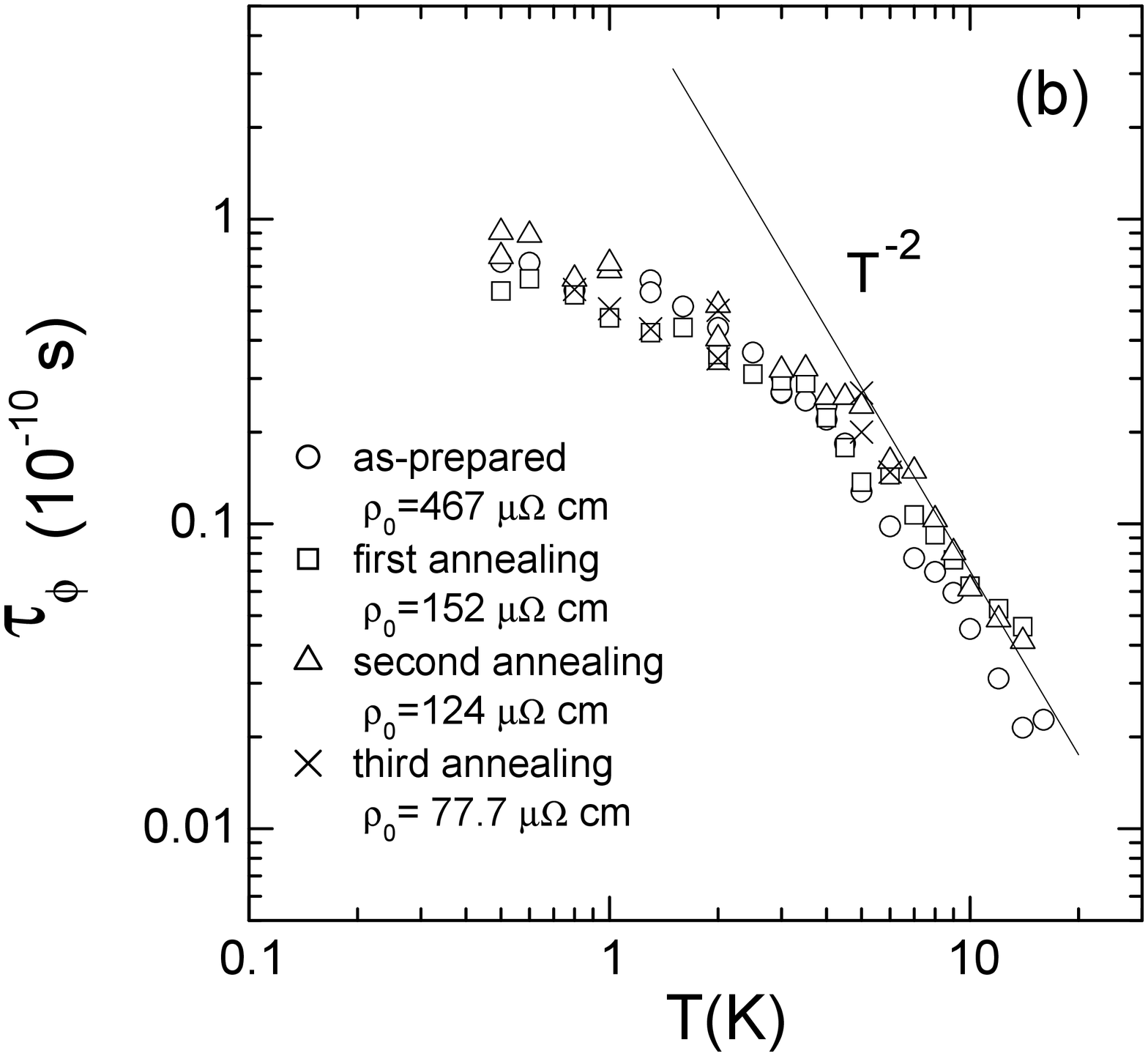}
\caption{Variation of $\tau_\phi$ with temperature for (a) a `moderately' disordered, and (b) a
`strongly' disordered, AuPd thick films before and after annealing. The solid line in (b) is a
guide to the eye. These figures were taken from Ref. \cite{Lin02}. \label{sf8}}
\end{center}
\end{figure}

The effect of annealing on 3D polycrystalline metals has been studied very recently. Lin {\it et
al.} \cite{Lin02} have performed systematic measurements of $\tau_\phi$ on several series of {\em
as-sputtered} and subsequently {\em annealed} AuPd and Sb thick films. Such controlled annealing
measurements \cite{such} are crucial for testing theoretical models of dephasing that invoke the
role of magnetic scattering and dynamical defects. Figure \ref{sf8}(a) shows a plot of the
variation of $\tau_\phi$ with $T$ for one of their as-prepared and subsequently annealed AuPd
thick film. This figure clearly indicates that $\tau_\phi$ is increased by annealing. At first
glance, it appears that this observation is easily explained. Suppose that annealing results in
the rearrangement of lattice atoms and a relaxation of grain boundaries, thereby making the film
less disordered. Because TLS are closely associated with defects in the microstructures, their
number concentration, $n_{\rm TLS}$, is therefore reduced by annealing. Assuming that dynamical
defects are effective scatterers as $T \rightarrow 0$, one can then understand Fig. \ref{sf8}(a)
in terms of a TLS picture, i.e. $\tau_\phi^0 \propto n_{\rm TLS}^{-1}$. However, it is impossible
to perform a quantitative comparison of the experiment with TLS theories \cite{Zawa99,Imry99}. The
difficulties lie in the facts that (i) the number concentration $n_{\rm TLS}$ in a particular
sample is unknown, (ii) the strength of coupling between conduction electrons and a TLS is poorly
understood, and (iii) the dynamical properties of real defects (impurities, grain boundaries,
etc.) are even less clear. Moreover, further measurements of Lin {\it et al.} indicate that the
nature of low-$T$ dephasing in polycrystalline metals is not so straightforward. They found that
the effect of annealing on $\tau_\phi$ is distinctly different in samples having much higher
resistivities.

In addition to the study of Fig. \ref{sf8}(a), Lin {\it et al.} \cite{Lin02} have also carried out
measurements on AuPd thick films containing much higher levels of disorder. Surprisingly, they
discovered that annealing has a {\em negligible} effect on $\tau_\phi$ in {\em strongly}
disordered AuPd thick films. Figure~\ref{sf8}(b) shows the variation of $\tau_\phi$ with
temperature for a strongly disordered AuPd thick film. This figure demonstrates that the values of
$\tau_\phi$ for the as-prepared and annealed samples are essentially the same, even though the
resistance is changed by the annealing by a factor of more than 6. The absence of an appreciable
annealing effect in this case implies that, in addition to the usual TLS addressed above, strongly
disordered films also contain other defects that cannot be readily cured by annealing. This
ineffectiveness of thermal annealing may suggest that there are two kinds of TLS. On the other
hand, it may suggest that, despite a large effort in this direction, no real defects of any nature
have dynamical properties which can explain the saturation of $\tau_\phi^0$ found in the
experiments \cite{Zawa99}. Inspection of the large discrepancy in Figs.~\ref{sf8}(a) and
\ref{sf8}(b) strongly indicates that low-$T$ dephasing is very sensitive to the microstructures in
the samples.

We return to the issue of magnetic scattering. The result of Figs. \ref{sf8}(a) and \ref{sf8}(b)
indicates that magnetic scattering should play a subdominant role, if any, in inducing the
saturation of $\tau_\phi^0$. The reasons are given as follows. (i) Suppose that there is a low
level of magnetic contamination in the as-sputtered film. Upon annealing, the magnetic impurity
concentration $n_m$ should be left unchanged. If the original saturation in the as-sputtered
sample is caused by spin-spin scattering, one should then expect the same value of $\tau_\phi^0$
($\propto n_m^{-1}$) after annealing. However, the result of Fig. \ref{sf8}(a) indicates an
increased $\tau_\phi^0$ with annealing, which is in disagreement with this assumption. (ii)
Blachly and Giordano \cite{ng95} have found that the Kondo effect is very {\em sensitive} to
disorder, namely that increasing disorder suppresses the Kondo effect. Along these lines, if the
original saturation of $\tau_\phi^0$ found in Fig.~\ref{sf8}(b) were really due to magnetic
scattering, one should then argue that annealing that suppresses disorder should enhance the Kondo
effect. Therefore, a decreased $\tau_\phi^0$ should be expected with annealing. Since the measured
$\tau_\phi^0$ does not change, even when the sample resistivity is reduced by a factor of more
than 6 by annealing, Fig.~\ref{sf8}(b) thus cannot be reconciled with a magnetic-scattering
scenario. This picture of a suppressed Kondo effect with increasing disorder is also incompatible
with the result for the moderately disordered film considered in Fig.~\ref{sf8}(a), where an
increased, instead of a decreased, $\tau_\phi^0$ is found after annealing. In short, systematic
annealing measurements in both thin \cite{Lin87b} and thick \cite{Lin02} films {\em cannot} be
reconciled with magnetic scattering being responsible for the saturation of $\tau_\phi^0$ at low
temperatures.

\subsection{The importance of three-dimensional structures}

It is worth noting that the saturation problem can be better addressed in 3D, rather than
lower-dimensional, structures. This is because of the increased contrast between the saturation
and the strong dependence of $\tau_{\rm i}(T)$ in 3D. As discussed in \S 2, the inelastic electron
scattering rate $\tau_{\rm i}^{-1} \approx \tau_{ep}^{-1} \propto T^p$, with $p \gtrsim 2$ in 3D.
Such a $T$ variation is much stronger than the dominating $p = 2/3$ in 1D and the $p =1$ in 2D.
For example, inspection of the solid line, which is drawn proportional to $T^{-2}$, in Fig.
\ref{sf8}(b) clearly reveals that the measured $\tau_\phi^0$ at 0.5 K is {\em already more than
one order of magnitude} lower than would be extrapolated from the measured $\tau_{ep}$ at a few
degrees Kelvin. Obviously, such a large discrepancy cannot simply be ascribed to experimental
uncertainty. (The increasing contrast between the saturation and the dependence of $\tau_{\rm
i}(T)$ with increasing sample dimensionality is already directly manifested in Fig. \ref{f1}.)
There is another advantage of using 3D structures in the studies of $\tau_\phi^0$. Compared with
the fabrication of narrow wires, the preparation of 3D samples usually does not require
sophisticated lithographic processing, thereby greatly minimizing any (magnetic) contamination
that might eventually act like a spin-flipper as $T \rightarrow 0$.

\section{Conclusion}

Over the years, the advances in our understanding of quantum-interference effects have made
feasible systematic and quantitative measurements of $\tau_\phi$ ($\tau_\phi^0$). Despite
extensive efforts in theoretical calculations and experimental measurements of $\tau_\phi^0$, our
current understanding of the microscopic origins for the zero-$T$ dephasing in {\em real}
conductors is still incomplete. Experimentally, carefully designed low-$T$ magneto-transport
measurements employing tailor-made structures, with sample specifics varying over a wide range of
disorder and dimensionality, would be highly desirable to help with discerning the underlying
physics of $\tau_\phi^0$. In addition to the systematic studies on high-disorder 3D
polycrystalline metals \cite{Lin01a,Lin02}, combined measurements of the electron energy exchange
rate, dephasing rate, and Aharonov-Bohm oscillations in the presence of a high magnetic field will
shed light on this issue \cite{Birge00}. Thus far, systematic measurements have ruled out electron
heating, microwave noise, and magnetic scattering as the dominant source for the saturation
behavior of $\tau_\phi^0$ observed in the experiments.

In addition to the case of disordered metals in the diffusive regime, a saturation of
$\tau_\phi^0$ has also been observed in semiconductor, diffusive and quasi-ballistic, quantum
wires, and ballistic dots \cite{Lin02r}. In many regards, the features of this saturation appear
reminiscent of that found in dirty metal wires. The saturated value of $\tau_\phi^0$ is typically
of similar order in semiconductor wires and dots, and is also of comparable magnitude to that
found in studies of dirty metal wires and films. The characteristic temperature for onset of the
saturation also varies widely in these structures---again reminiscent of the behavior found in
dirty mesoscopic systems. It is intriguing, and deserves serious investigation, why semiconductor
quantum structures and dirty metals reveal similar saturation behavior of $\tau_\phi^0$.

Another important issue revealed in a number of studies is a sensitivity of the electron dephasing
to the microscopic quality of disorder. For instance, Ovadyahu \cite{Ovadyahu01} has measured
$\tau_\phi$ at low temperatures in diffusive In$_2$O$_{3-x}$ and In$_2$O$_{3-x}$:Au thin films. He
found that, although the Au doping is only $\lesssim$ 3\% in In$_2$O$_{3-x}$:Au thin films, the
behavior of the dephasing time in these two materials could be {\em significantly different}. Bird
and co-workers \cite{Bird98a} have studied semiconductor quantum dots and found that their
$\tau_\phi$ can show significant dot-to-dot variations, in samples realized in materials with
similar mobilities. These measurements reflect a critical sensitivity of the dephasing processes
to disorder. These experiments clearly suggest that $\tau_\phi^0$ is not only dependent on the
{\em total} level of disorder, but are also very sensitive to the microscopic {\em quality} of the
disorder. This can be particularly crucial for mesoscopic devices, whose disorder profile is know
to be highly sample specific. This is a key point that needs to be taken into consideration in
future theories of electron dephasing times.

\section{Acknowledgements}

This work was supported by Taiwan National Science Council through Grant No. NSC90-2112-M-009-037.


\begin{references}
\bibitem{such} We notice that annealing experiment may not be performed on metallic
glasses, because thermal annealing can cause crystallization of amorphous samples.
\end{references}
\end{document}